\documentclass{aastex}          
\usepackage{spr-astr-addons}    
\usepackage{epsfig}
\usepackage{graphicx}
\usepackage{color}

\begin{document}
%
\title{Transverse oscillations in solar spicules induced by propagating Alfv\'{e}nic pulses}

\author{H.~Ebadi\altaffilmark{1}}
\affil{Astrophysics Department, Physics Faculty,
University of Tabriz, Tabriz, Iran\\
e-mail: \textcolor{blue}{hosseinebadi@tabrizu.ac.ir}}
\and
\author{M.~Hosseinpour}
\affil{Plasma Physics Department, Physics Faculty,
University of Tabriz, Tabriz, Iran}
\and
\author{Z.~Fazel}
\affil{Astrophysics Department, Physics Faculty,
University of Tabriz, Tabriz, Iran}

\altaffiltext{1}{Research Institute for Astronomy and Astrophysics of Maragha,
Maragha 55134-441, Iran.}

\begin{abstract}
The excitation of Alfv\'{e}nic waves in the solar spicules due to the localized Alfv\'{e}nic pulse is investigated.
A set of incompressible MHD equations in two dimensional $x-z$ plane with steady flows and sheared
magnetic fields is solved. Stratification due to gravity and transition region between chromosphere and corona are taken into account.
An initially localized Alfv\'{e}nic pulse launched below the transition region can
penetrate from transition region into the corona.
We show that the period of transversal oscillations is in agreement with those observed in spicules.
Moreover, it is found that the excited Alfv\'{e}nic waves spread during propagation along the spicule length, and
suffer efficient damping of the oscillations amplitude. The damping time of transverse oscillations elongated with decrease in $k_{b}$
values.

\end{abstract}


\section{Introduction}
\label{sec:intro}
Spicules are one of the most fundamental components of the solar chromosphere.
They are seen in chromospheric spectral lines at the solar limb at speeds of about
$20-25$~km~s$^{-1}$ propagating from the chromosphere into the solar corona \citep{Tem2009}. Their
diameter and length varies from spicule to spicule having the values from $400$~km
to $1500$~km and from $5000$~km to $9000$~km, respectively. The typical lifetime of them is $5-15$ min.
The typical electron density at heights where the spicules are observed
is approximately $3.5\times10^{16}-2\times10^{17}$ m$^{-3}$, and their temperatures are estimated as $5000-8000$ K \citep{bec68, ster2000}.
\citet{Kukh2006, Tem2007} by analyzing the height series of $H\alpha$ spectra
in solar limb spicules observed their transverse oscillations with the estimated period of $20-55$ and $75-110$ s.
More recently, \citet{Ebadi2012a} based on \emph{Hinode}/SOT observations
estimated the oscillation period of spicule axis around $180$ s.
\\Despite the large body of theoretical and observational works devoted to the spicules,
their ejection mechanism is not clear yet.
In other words, observations with high spatial resolutions are needed to distinguish their origin.
The Alfv\'{e}n waves are usual candidates for energy transport from the lower layers of
the solar atmosphere to the corona \citep{De2007, Tsiklauri2002}.
\citet{Hollweg1982} showed that the Alfv\'{e}n waves may be nonlinearly
coupled to fast magnetoacoustic shocks, which may lead to spicule formation.
\citet{Cargill1997} performed the numerical simulations of the propagation
of Alfv\'{e}nic pulses in two dimensional magnetic field geometries.
They concluded that for an Alfv\'{e}nic pulse the time at which different parts of
the pulse emerge into the corona depends on the plasma density and magnetic field properties.
Moreover, they discussed that this mechanism can interpret spicule ejection forced through the
transition region.
\citet{Kudoh1999} used the random nonlinear Alfv\'{e}nic pulses and
concluded that the transition region lifted up to more than $\sim5000$ km (i.e. the spicule produced).
\citet{Del2005} studied the propagation and evolution of Alfv\'{e}nic pulses through coronal arcades.
Their results showed a strong spreading of the initially localized pulses due to variations of Alfv\'{e}n
speed with height and efficient damping of oscillations. In addition, they concluded that the background
density and Alfv\'{e}n speed should be considered as key ingredients in models.
\citet{Tem2010} studied the upward propagation of a velocity pulse launched initially below the transition region.
The pulse quickly steepens to a shock which may form the spicules.
They concluded that such a model may explain speed, width, and heights of classical spicules.
\citet{Rae1982} illustrated that impulsive motions may propagate along an intense tube.
Information (in the form of a wave front) propagates at the subsonic and subAlfv\'{e}nic tube speed.
As the wave front of an impulsively generated disturbance propagates along the tube,
it trails behind it a wake, which oscillates at cut-off frequency.
The significance of cut-off frequency depends upon what are the circumstances creating the motions.
If disturbances are impulsively generated, then wave propagation occurs and the existence of
cut-off frequency is manifest in the creation of an oscillatory wake.
\citet{Hasan1999} examined the generation of waves in vertical flux tubes
in the magnetic network through the impulsive excitation by granules.
The buffering action of a granule on a flux tube at a certain level
excites a pulse that travels away from the source region with tube speed.
After the passage of the pulse, the atmosphere oscillates at the cut-off
period of the mode, with amplitude that slowly decays in time.
\citet{Tem2008} concluded that photospheric granulation may excite
transverse pulses in anchored vertical magnetic flux tubes.
The pulses propagate upward along the tubes while oscillating wake
is formed behind the wave front in a stratified atmosphere.
The pulses carry almost all the energy of initial perturbations,
while the energy in wake oscillations is much smaller.
\\In the present work we perform two-dimensional simulations of MHD equations and show that the Alfv\'{e}nic pulse
which is launched below the transition region can excite Alfv\'{e}n waves and reach the height where spicules are observed.
Section $2$ gives the basic equations and theoretical model. In section $3$ numerical
 results are presented and discussed, and a brief summary is followed in section $4$.

\section{Theoretical modeling}
\label{sec:theory}
\begin{figure}
\centering
\includegraphics[width=8cm]{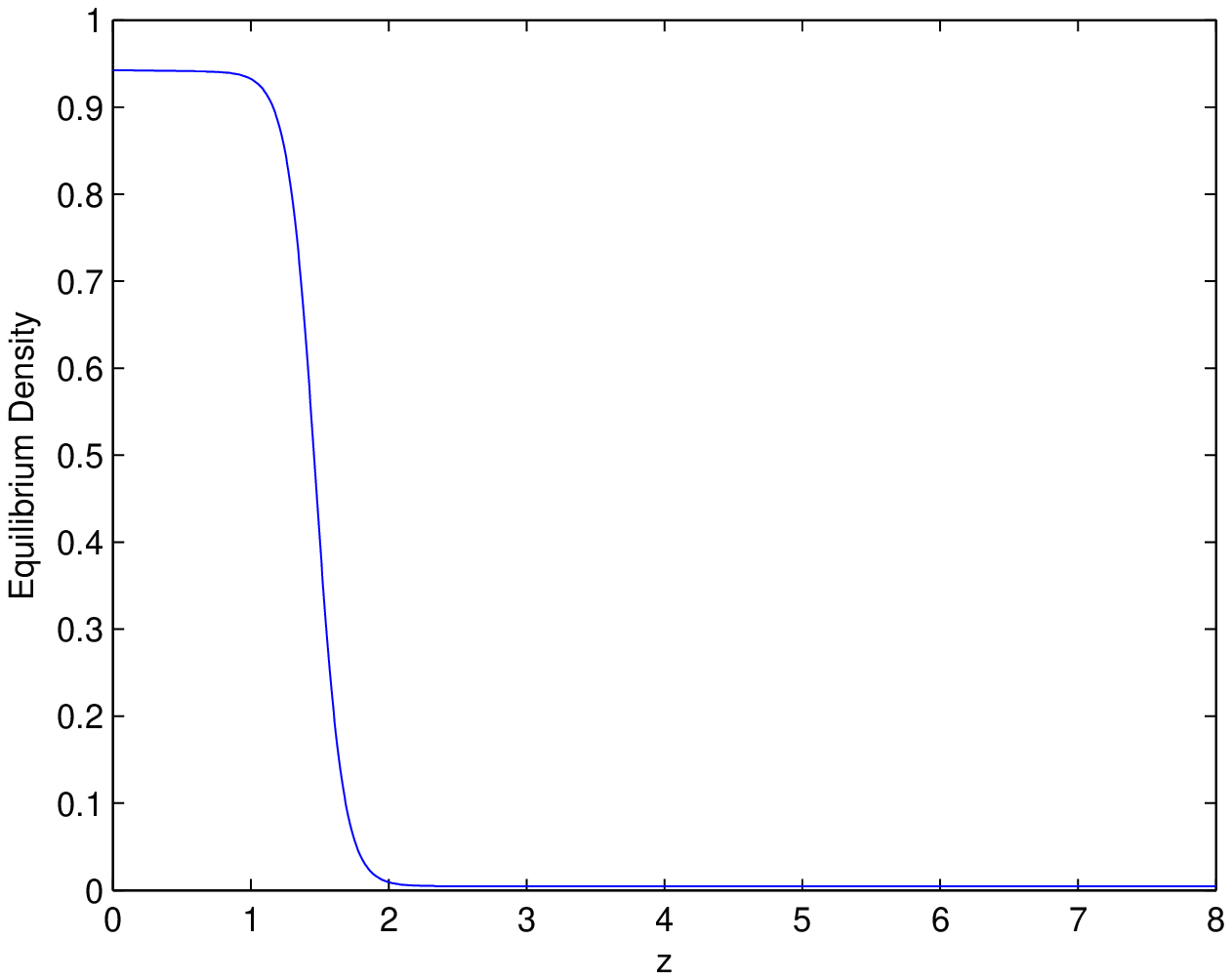}
\includegraphics[width=8cm]{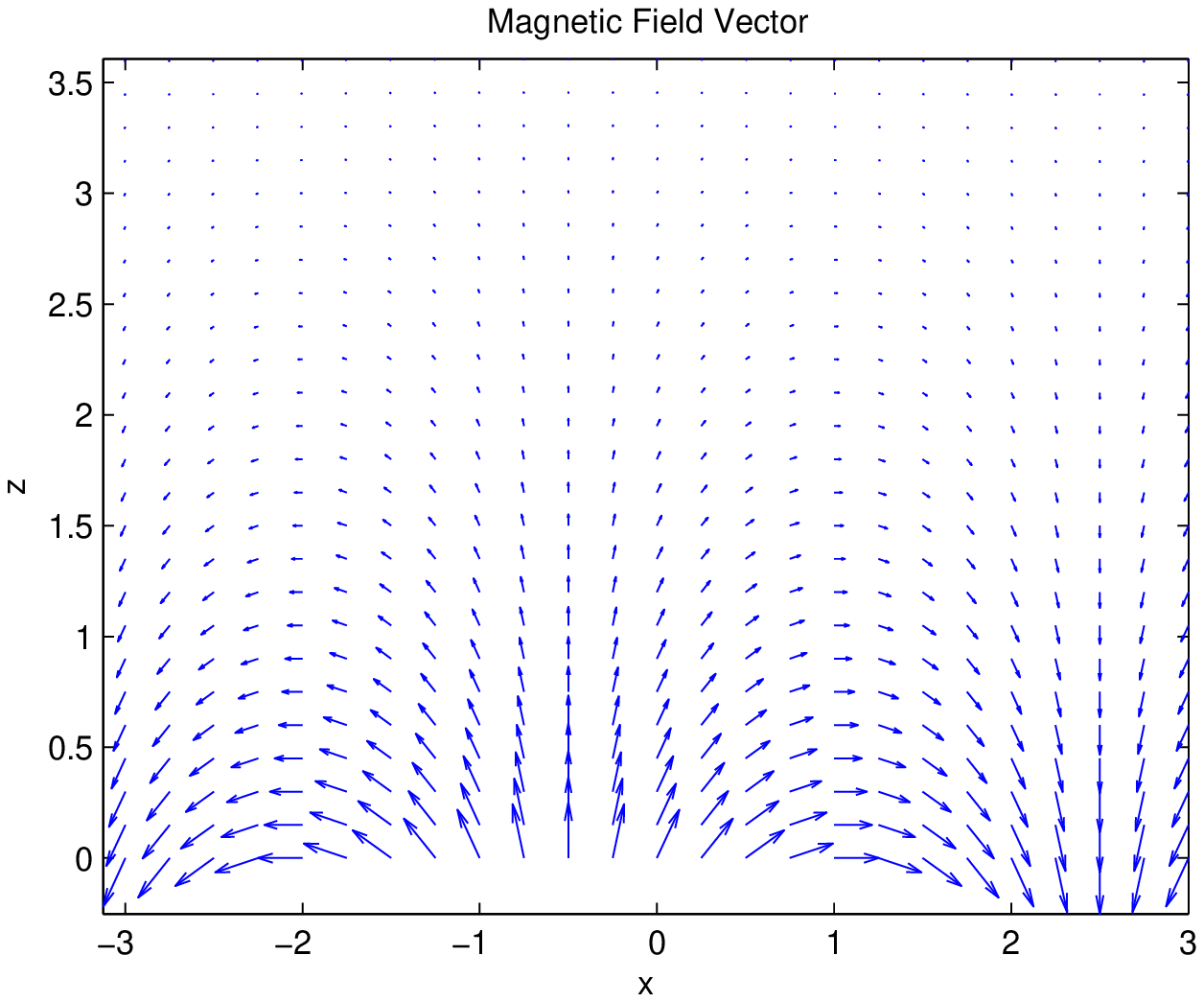}
\caption{(Color online) Dimensionless equilibrium density and total magnetic field plots with respect
to dimensionless $x$, and $z$ space coordinates ($k_{b}=\pi/3$).  \label{fig1}}
\end{figure}

We keep effects of stratification due to
gravity in $2$D x-z plane in the presence of steady flow and shear field.
Since the excitation of incompressible Alfv\'{e}n waves due to Alfv\'{e}nic pulses is the
main goal of this work, so the continuity and energy equations are not considered.
The ideal MHD equations used are as follows:
\begin{equation}
\label{eq:momentum} \rho(\frac{\partial \mathbf{v}}{\partial t}+
\mathbf{v} \cdot \nabla\mathbf{v}) = -\nabla p + \rho
\mathbf{g}+ \frac{1}{\mu}(\nabla \times \mathbf{B})\times
\mathbf{B},
\end{equation}
\begin{equation}
\label{eq:induction} \frac{\partial \mathbf{B}}{\partial t} = \nabla
\times(\mathbf{v} \times \mathbf{B}),
\end{equation}

\begin{equation}
\label{eq:state} p = \frac{\rho RT}{\mu}.
\end{equation}
Here, $\rho$ is the plasma density; $T$ is the plasma temperature in Kelvin; $R$ is the universal gas constant, and $\mu$ is the mean molecular weight.
We assume that the spicules are highly dynamic with speeds that are
significant fractions of the Alfv\'{e}n speed. Perturbations are assumed independent of y, i.e.:
\begin{eqnarray}
\label{eq:perv}
  \textbf{v} &=& v_{0} \hat{k} + v_{y}(x,z,t) \hat{j} \nonumber\\
  \textbf{B} &=& B_{0x}(x,z) \hat{i}+B_{0z}(x,z) \hat{k} + b_{y}(x,z,t) \hat{j},
\end{eqnarray}
and the equilibrium sheared magnetic field is two-dimensional and divergence-free as \citep{Del2005,Tem2010}:

\begin{eqnarray}
\label{eq:shear field}
 B_{0x}(x,z) &=& B_{0}e^{-k_{b}z} \cos[k_{b}(x-a)] \nonumber\\
 B_{0z}(x,z) &=& -B_{0}e^{-k_{b}z} \sin[k_{b}(x-a)] ,
\end{eqnarray}
with $a$ as the spicule radius.
Since the equilibrium magnetic field is force-free, therefore the pressure gradient is balanced by the gravity force, which is assumed
to be \textbf{g}=-g $\hat{k}$ via this equation:

\begin{equation}
\label{eq:balance}
 -\nabla p_{0}(z) + \rho_{0}(z) \textbf{g}=0,
\end{equation}
and the pressure in an equilibrium state is:
\begin{equation}
\label{eq:presse}
 p_{0}(z)= p_{0}~\exp\left(-\int^{z}_{z_{r}}\frac{dz'}{\Lambda(z')}\right).
\end{equation}
The density profile is in the form of:

\begin{equation}
\label{eq:density}
 \rho_{0}(z)= \frac{\rho_{0}T_{0}}{T_{0}(z)}~\exp\left(-\int^{z}_{z_{r}}\frac{dz'}{\Lambda(z')}\right),
\end{equation}
with
\begin{equation}
\label{eq:scale}
 \Lambda(z)= \frac{RT_{0}(z)}{\mu g},
\end{equation}
where the temperature profile is taken here as a smoothed step function, i.e.:
\begin{equation}
\label{eq:temp}
 T_{0}(z)= \frac{1}{2}T_{c}\left [1+d_{t}+(1-d_{t})\tanh(\frac{z-z_{t}}{z_{\omega}})\right],
\end{equation}
here $d_{t}=T_{ch}/T_{c}$ with $T_{ch}$ denoting the chromosphere temperature at its lower part.
The symbol $T_{c}$ corresponds to the temperature of the solar corona that is separated from the chromosphere
by the transition region, which has the width of $z_{w}=200~km$, and is located at the $z_{t}=2000 ~km$ above the solar surface.
We put $T_{ch}=15\times10^{3}~K$ and $T_{c}=3\times10^{6}~K$.
\\Figure~\ref{fig1} shows the equilibrium density and magnetic field with respect to $x$, and
$z$ (all quantities in these plots are non-dimensional).

The linearized dimensionless MHD equations with these assumptions are:

\begin{equation}
\label{eq:veloy}
 \frac{\partial v_{y}}{\partial t} + v_{0}\frac{\partial v_{y}}{\partial z}
 = \frac{1}{\rho_{0}(z)}\left[B_{0x}(x,z)\frac{\partial b_{y}}{\partial x}+B_{0z}(x,z)\frac{\partial b_{y}}{\partial z}\right],
\end{equation}

\begin{equation}
\label{eq:mag}
 \frac{\partial b_{y}}{\partial t} + v_{0}\frac{\partial b_{y}}{\partial z}
 = B_{0x}(x,z)\frac{\partial v_{y}}{\partial x}+B_{0z}(x,z)\frac{\partial v_{y}}{\partial z},
\end{equation}

where the densities, velocities, the magnetic field, time and space coordinates are normalized to
$\rho_{\rm 0}$ (the plasma density at dimensionless $z=6$), V$_{A0}\equiv B_{\rm 0}/\sqrt{\mu_{0} \rho_{\rm 0}}$,
$B_{\rm 0}$, $\tau$ (the time scale of transit Alfv\'{e}n time defined as $\tau=a/V_{A0}$), $a$ (spicule radios), respectively.
Also the gravity acceleration is normalized to $a/\tau^{2}$.
\\Eqs.~\ref{eq:veloy}, and ~\ref{eq:mag} should be solved under following
initial conditions:

\begin{eqnarray}
\label{eq:icv}
  v_{y}(x,z,t=0) &=& A_{v}\exp \left [-\frac{(x-x_{0})^{2}+(z-z_{0})^{2}}{w^{2}}\right] \nonumber\\
  b_{y}(x,z,t=0) &=& A_{b} \sin(\pi x)\sin(\pi z) ,
\end{eqnarray}

where the initial magnetic perturbation amplitude is set to be $A_{b}=10^{-7}$,
and ($x_{0}$, $z_{0}$) determine the center of an initial pulse where it has the maximum speed.

\section{Numerical results and discussion}
To solve the coupled Eqs.~\ref{eq:veloy}, and~\ref{eq:mag} numerically,
the finite difference and the Fourth-Order Runge-Kutta methods
are used to take the space and time derivatives, respectively.
We set the number of mesh-grid points as~$256\times256$.
In addition, the time step is chosen as $0.001$, and the system length in the $x$ and $z$ dimensions
(simulation box sizes) are set to be ($0$,$2$) and ($0$,$8$).
The parameters in spicule environment are as follows:
$a$ (spicule radios)=1000 km,
$\omega=0.3a=300 km$ (the width of Gaussian packet), L=8000 km (Spicule length), $v_{0}=25 km/s$,
$n_{e}=11.5\times10^{16} m^{-3}$, $B_{0}=1.2\times10^{-3}~Tesla$,
 $T_{0}=14~000~K$, $g=272~m s^{-2}$, $R=8300~m^{2}s^{-1}k^{-1}$ (universal gas constant),
$V_{A0}=77.5~km/s$, $\mu=0.6$, $\tau=13$ s ,
$\rho_{0}=1.9\times10^{-10}~kg m^{-3}$, $p_{0}=3.7\times10^{-2}~N m^{-2}$,
 $\mu_{0}=4\pi \times10^{-7}~TmA^{-1}$, $z_{r}=6000~km$ (reference height), $z_{w}=200~km$,
$z_{t}=2000~km$, $A_{v}=7.75~km/s$, $x_{0}=1$, $z_{0}=0.5$ \citep{Ebadi2012b,Ebadi2013}.

\begin{figure}
\centering
\includegraphics[width=8cm]{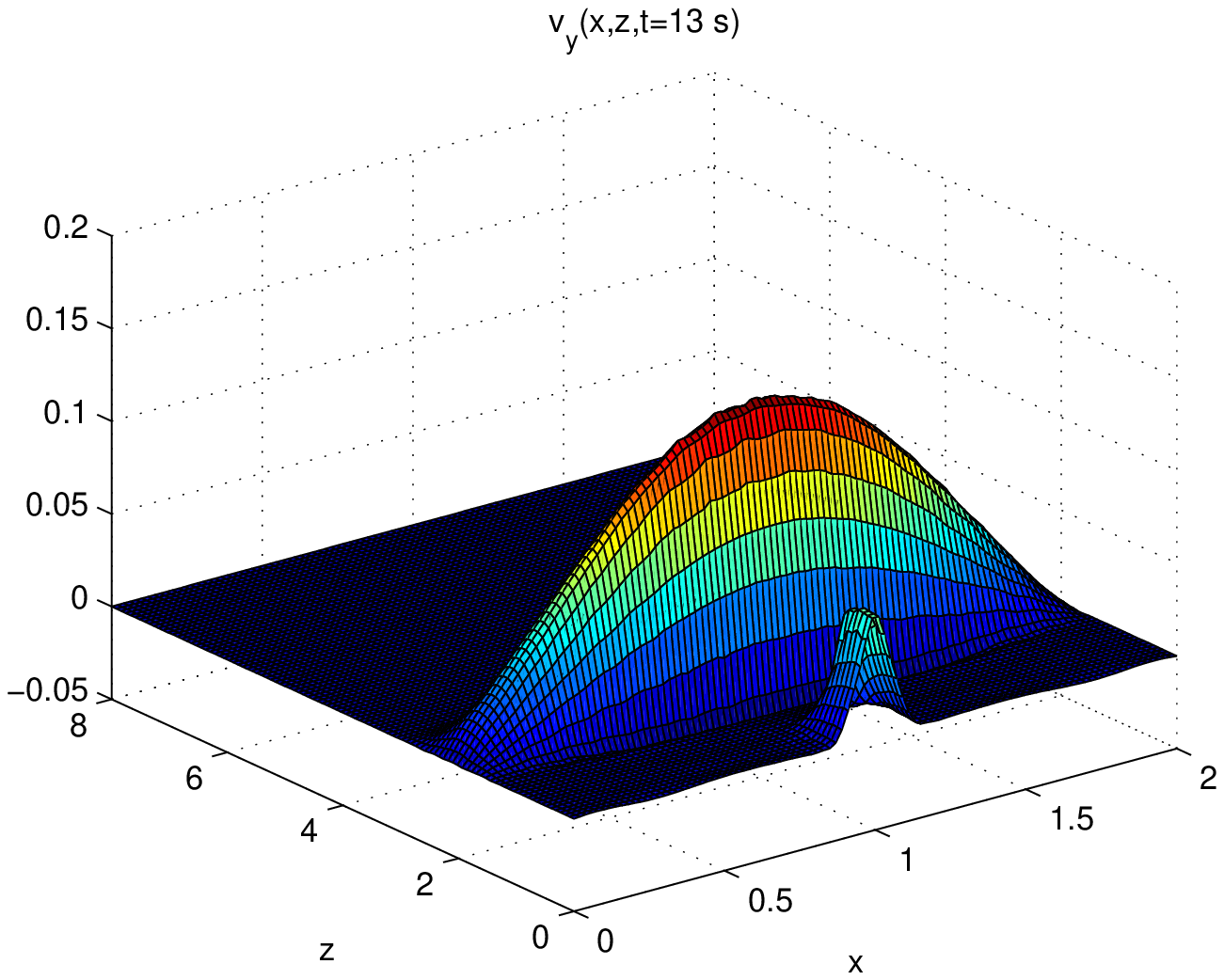}
\includegraphics[width=8cm]{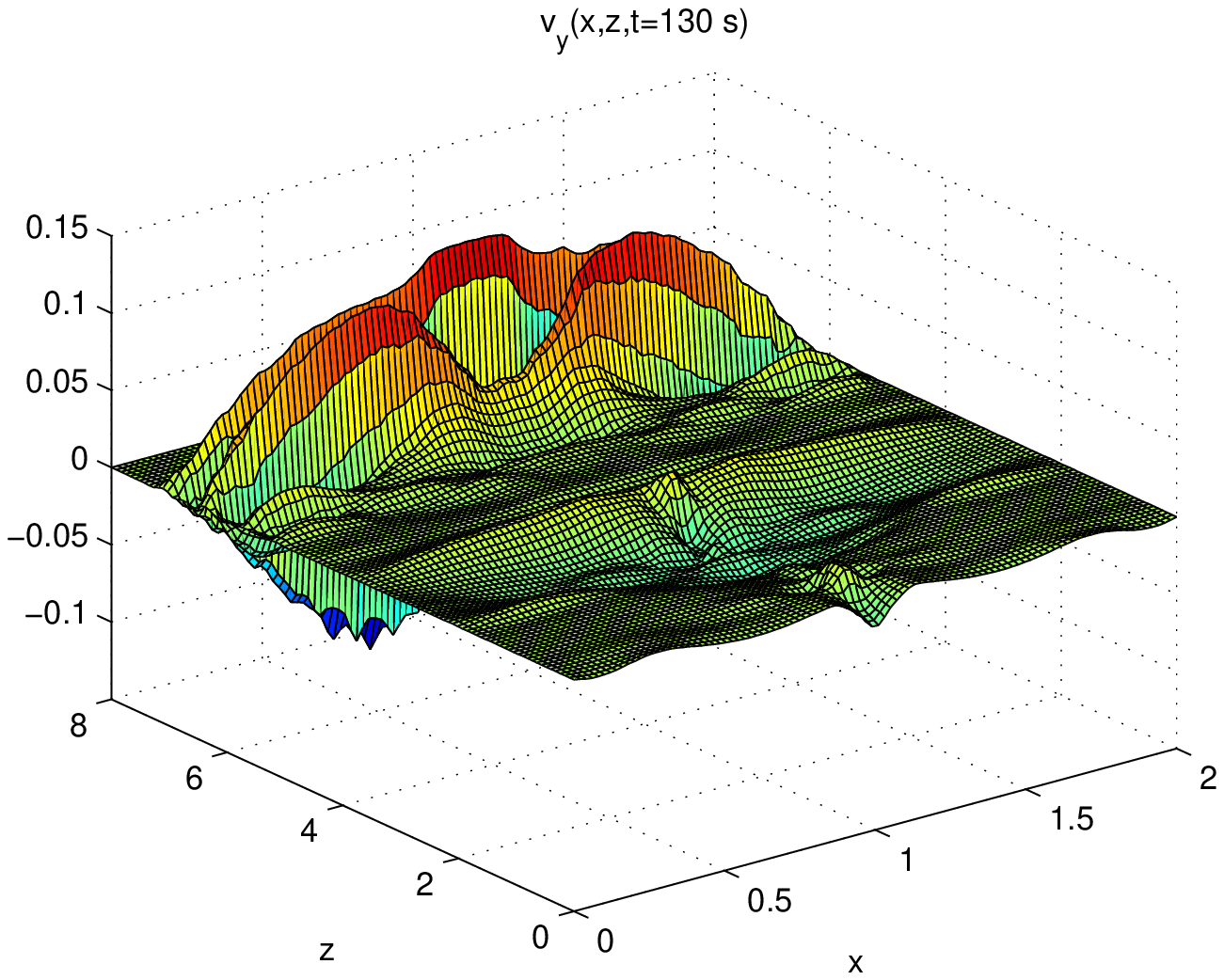}
\includegraphics[width=8cm]{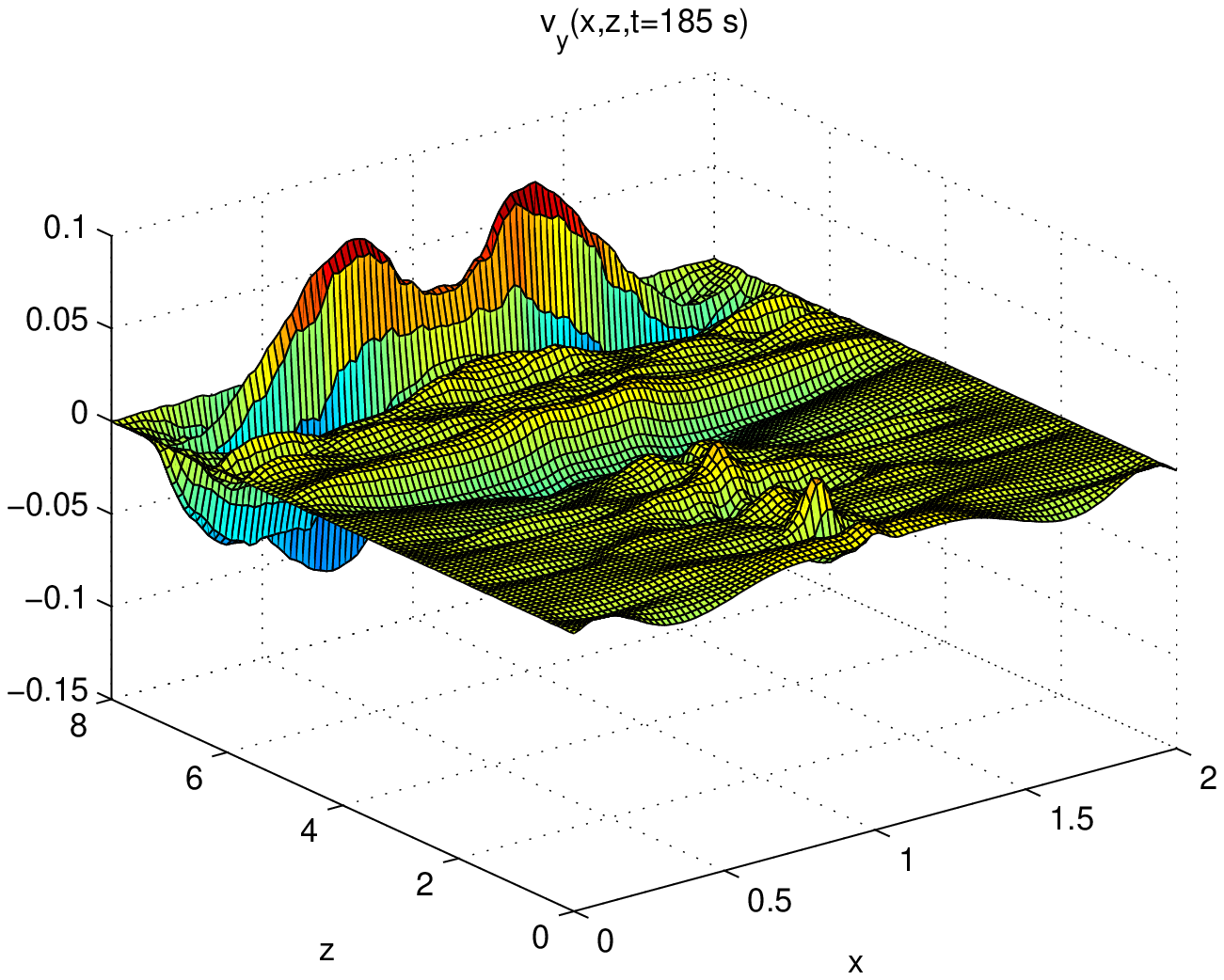}
\caption{(Color online) The $3D$ plots of the transversal component of the perturbed velocity with respect to
$x$, $z$ in $t=13 s$, $t=130 s$, and $t=185 s$ for $k_{b}=\pi/3$. \label{fig2}}
\end{figure}

Figure~\ref{fig2} shows $3D$ plots of the perturbed velocity with respect to
$x$, $z$ at $t=13 s$, $t=130 s$, and $t=185 s$ for $k_{b}=\pi/3$. The initial pulse amplitude is $7.75$ km/s
and is located in $x=1000$ km, and $z=500$ km. It shows that the initially localized
Alfv\'{e}nic pulse, which is launched below the transition region, propagates upward. We prefer to show the
$3D$ plots of the perturbed velocity in the case of $k_{b}=\pi/3$. Putting $k_{b}=\pi/3$ in the equilibrium magnetic
field leads to a pulse which propagates slowly regarding the total propagation distance. So that, analyzing the temporal behavior
of the pulse becomes easy.
As seen, the initially localized pulse during propagation suffers significant spreading along the spicule length
due to the variations of Alfv\'{e}n speed with height. This leads to an efficient damping of the oscillations amplitude (phase mixing).

\begin{figure}
\centering
\includegraphics[width=8cm]{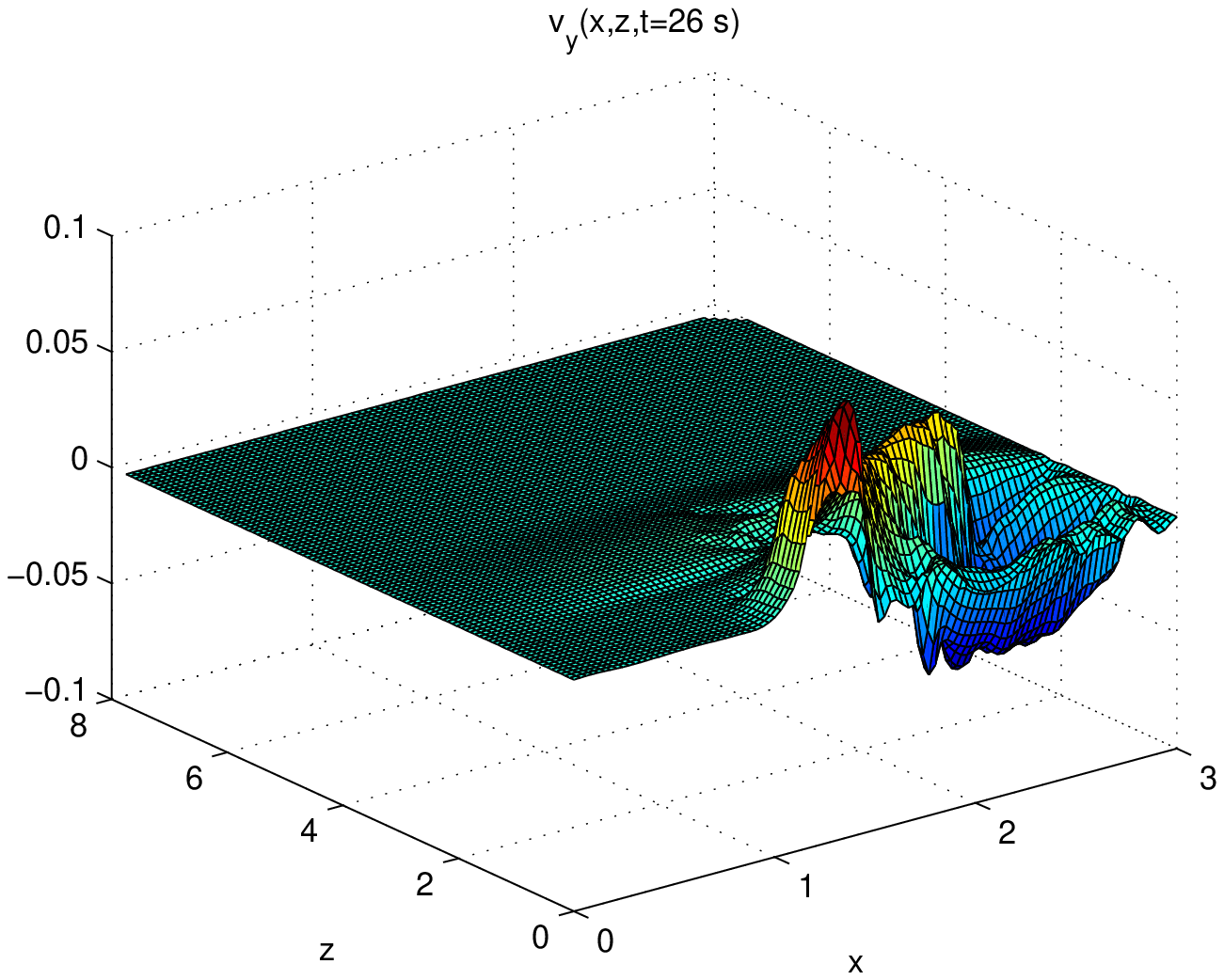}
\includegraphics[width=8cm]{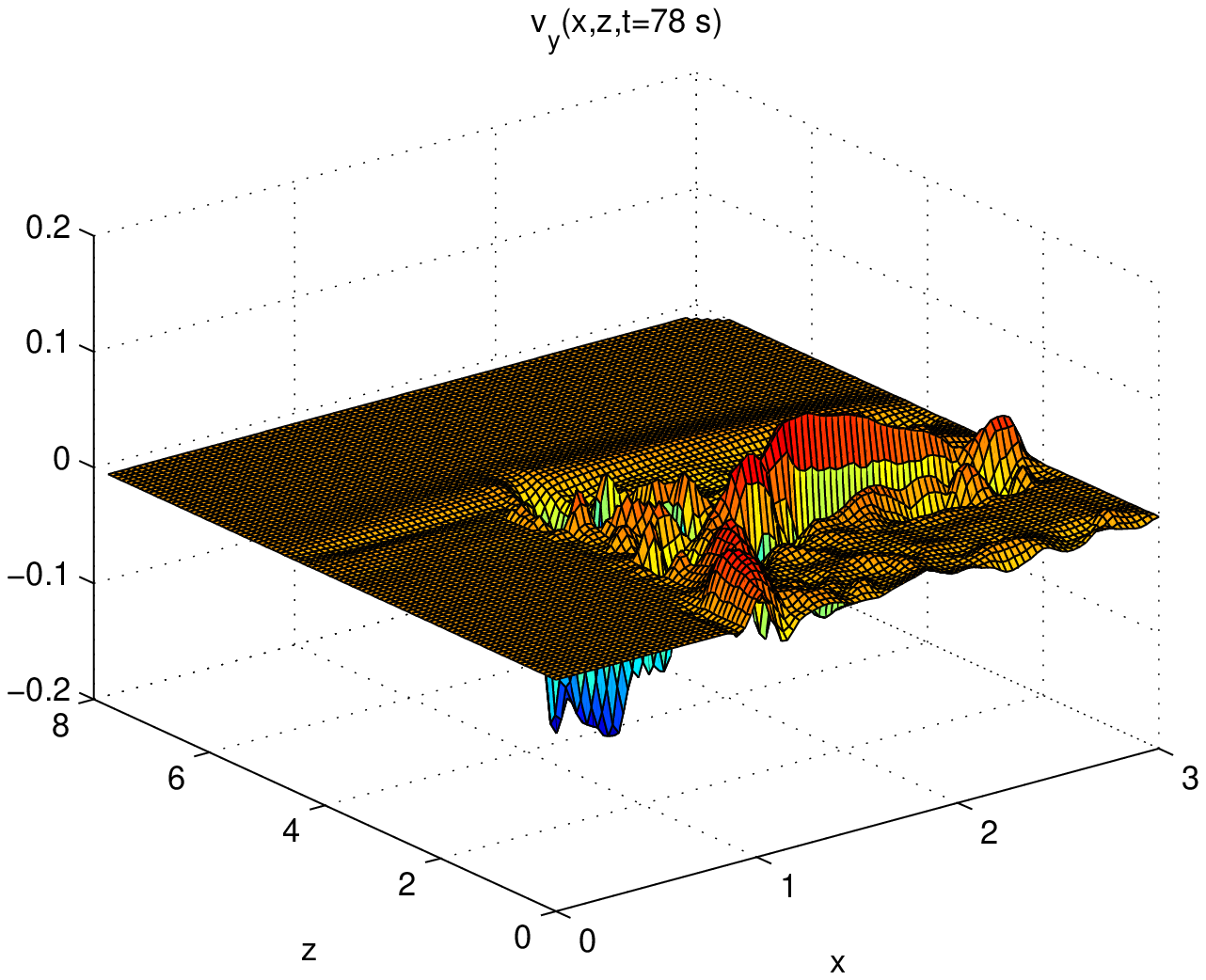}
\includegraphics[width=8cm]{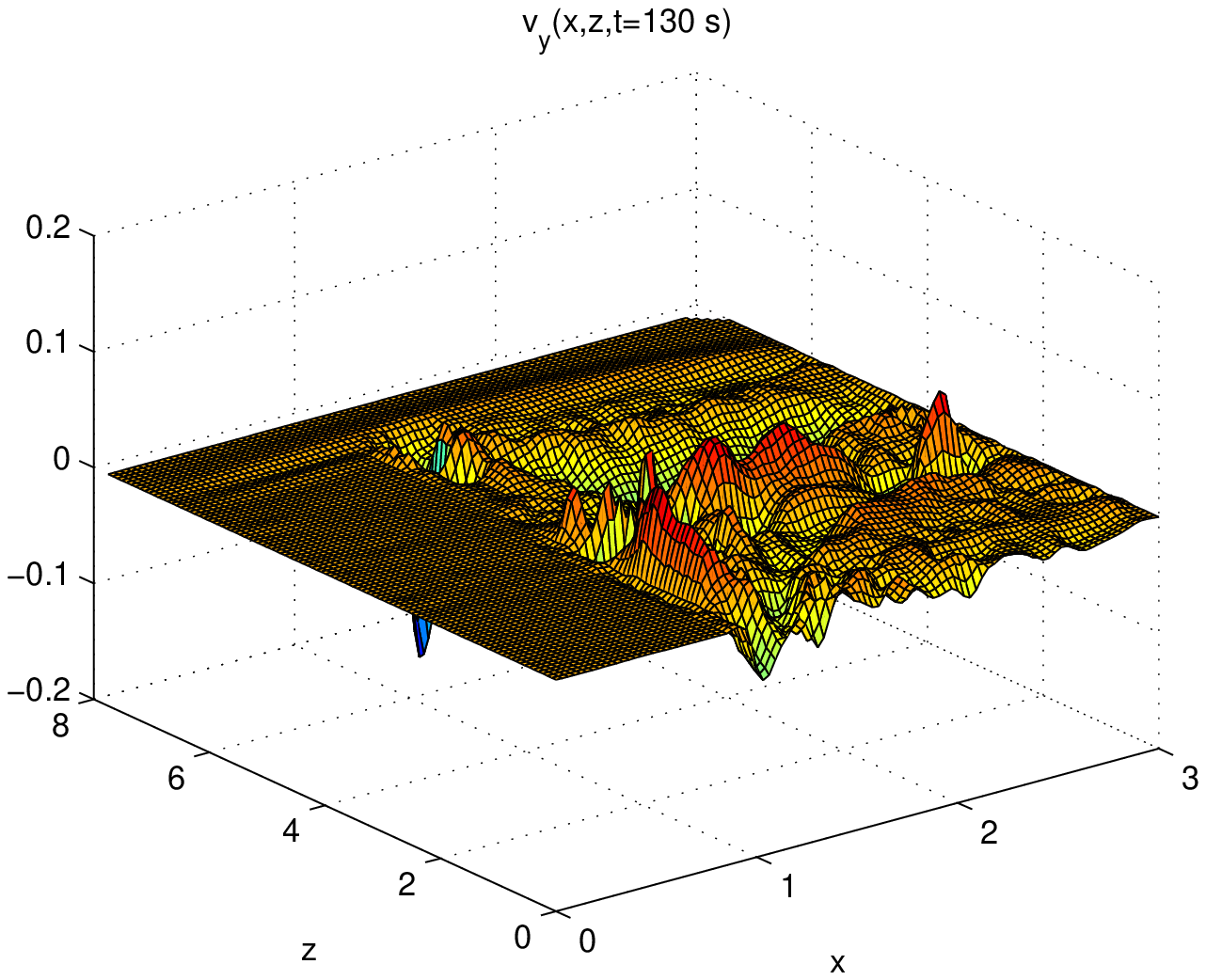}
\caption{(Color online) The $3D$ plots of the transversal component of the perturbed velocity with respect to
$x$, $z$ in $t=26 s$, $t=78 s$, and $t=130 s$ for $k_{b}=\pi/3$. \label{fig3}}
\end{figure}

Similar to figure~\ref{fig2}, in figure~\ref{fig3} the same plots of the perturbed velocity with respect to  
$x$, $z$ at $t=26 s$, $t=78 s$, and $t=130 s$ for $k_{b}=\pi/3$ are presented, 
but the initial pulse is now located at $x=2500$ km, and $z=500$ km. 
Obviously, the similar pattern of upward propagation of the initially localized Alfv\'{e}nic pulses repeated here.

\begin{figure*}[H]
\centering
\includegraphics[width=14cm]{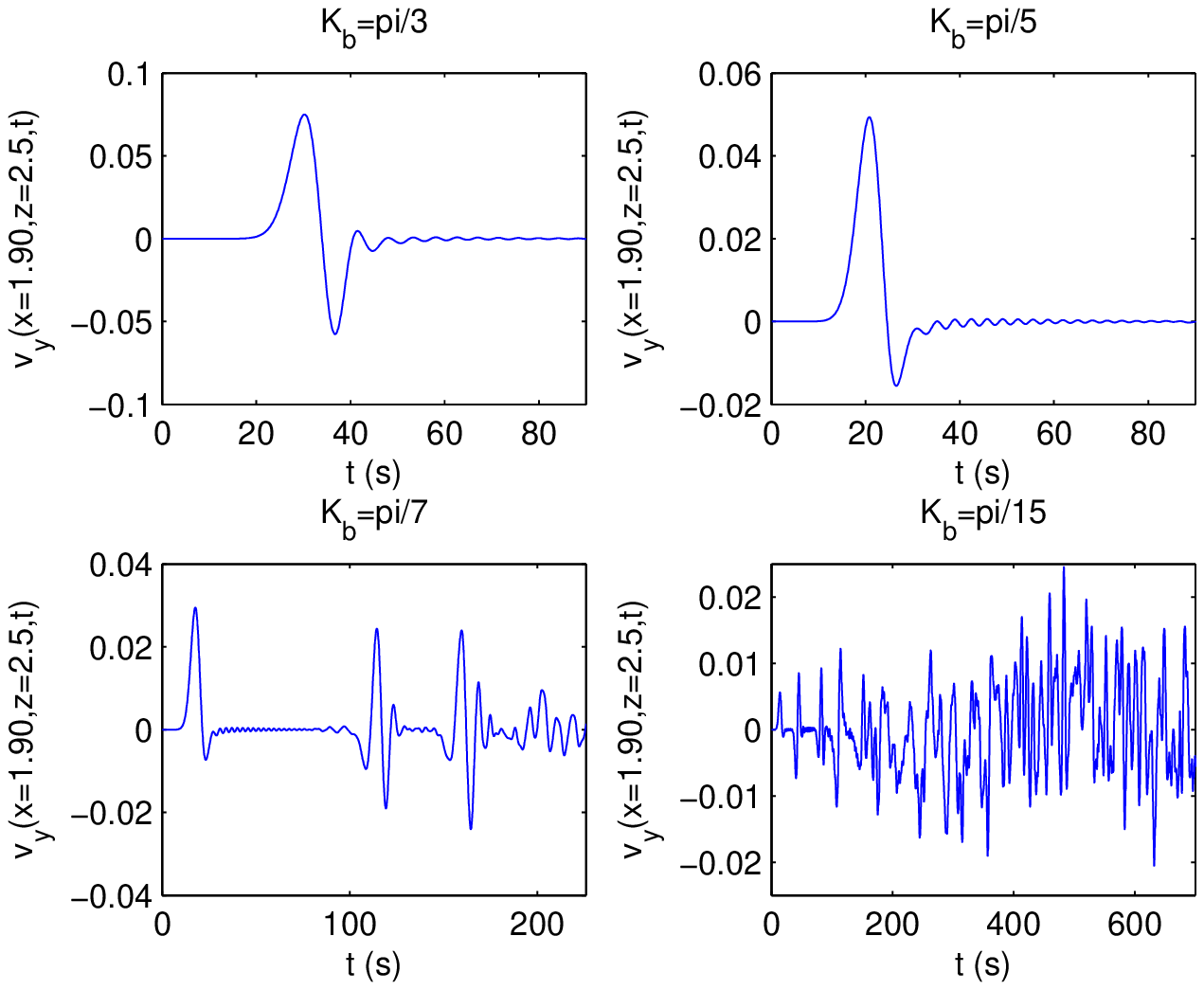}
\caption{(Color online) The time variations of transversal component of perturbed velocity in
detection point ($x=1900$ km, and $z=2500$ km) for different values of $k_{b}$. \label{fig4}}
\end{figure*}

Figure~\ref{fig4} illustrates variations of transversal component of the perturbed velocity with
time in $x=1900$ km, and $z=2500$ km for different values of $k_{b}$. The initial pulse may excite
the wake in stratified atmosphere which oscillates at cut-off frequency.
In the top-left panel ($k_{b}=\pi/3$) of figure~\ref{fig4},
the oscillatory wake behind the initial pulse is presented. The period of the oscillatory
wake corresponds to the period of transverse
oscillations of spicules observed by \citet{Tem2007}.
The same pattern is appeared in other panels but with different
periodicity values.
An another interesting result, which can be seen in different panels of figure~\ref{fig4} is
that the damping time of oscillations is elongated towards the smaller $k_{b}$ values.
The phase mixing and stratification due to gravity are the responsible mechanisms in damping of
the oscillations, and their efficiency reduce in small $k_{b}$ values \citep{Ebadi2012b,Ebadi2013}.
We have not found any observational
signature of this result in the literature (thanks to De Pontieu, B. for his useful comment about this point).
It is, in fact, a very difficult measurement.

\section{Conclusion}
\label{sec:concl}
The initially localized Alfv\'{e}nic pulse propagation in a medium with steady flows and
sheared magnetic field is investigated. We take into account the stratification due to
gravity and the transition region between chromosphere and corona. An initially localized
pulse can excite the Alfv\'{e}nic waves in the medium.
It is shown that these waves can penetrate from the transition region into the corona.
The period of transverse oscillations (wake) that are induced in the medium due to
the propagation of the Alfv\'{e}nic waves are in agreement with those observed in spicules.
Moreover, It is discussed that the excited Alfv\'{e}nic waves during propagation along the spicule length
suffer efficient damping of the oscillations amplitude. This can be related to the variations of Alfv\'{e}n speed with height
(phase mixing) and gravitational stratification. It is interesting that the damping time of oscillations
is elongated with the smaller $k_{b}$ values.
In other words, the phase mixing and stratification efficiency reduce in small $k_{b}$ values.

\acknowledgments
This work has been supported financially by the Research Institute for Astronomy and
Astrophysics of Maragha (RIAAM) under research project No. 1/2782-38, Maragha, Iran.

\makeatletter
\let\clear@thebibliography@page=\relax
\makeatother

\end{document}